\documentclass[titlepage,12pt]{article}
\usepackage{amssymb,amsmath,amsfonts}
\usepackage[utf8]{inputenc}
\usepackage{setspace}
\usepackage{epsfig}
\usepackage{graphicx}
\usepackage{cite}
\usepackage{caption}
\usepackage{enumitem}

\makeatletter
\def\@sect#1#2#3#4#5#6[#7]#8{\ifnum #2>\c@secnumdepth
  \def\@svsec{}\else
  \refstepcounter{#1}\edef\@svsec{\csname the#1\endcsname.\hskip0.5em}\fi
  \@tempskipa #5\relax
  \ifdim \@tempskipa>\z@
    \begingroup
      #6\relax
      \@hangfrom{\hskip #3\relax\@svsec}{\interlinepenalty \@M #8\par}%
    \endgroup
    \csname #1mark\endcsname{#7}\addcontentsline
      {toc}{#1}{\ifnum #2>\c@secnumdepth \else
        \protect\numberline{\csname the#1\endcsname}\fi #7}%
  \else
    \def\@svsechd{#6\hskip #3\@svsec #8\csname #1mark\endcsname
      {#7}\addcontentsline{toc}{#1}{\ifnum #2>\c@secnumdepth \else
        \protect\numberline{\csname the#1\endcsname}\fi #7}}%
  \fi \@xsect{#5}}

\newcommand{\D}[1]{{\cal D}_{#1}}

\begin{document}
\begin{titlepage}
  \begin{flushright}
TTK-14-04 
    \end{flushright}
\vspace{0.01cm}
\begin{center}
{\LARGE {\bf Decays of polarized top quarks to lepton, neutrino and
    jets at NLO QCD} } \\
\vspace{1.5cm}
\large{\bf W. Bernreuther\, \footnote{\tt
    breuther@physik.rwth-aachen.de},
     P.~Gonz\'alez\,\footnote{{\small Present address:
           Department of Radiation Oncology,
  Netherlands Cancer Institute, 1066 CX Amsterdam, NL.}} and C. Mellein\,\footnote{\tt
    mellein@physik.rwth-aachen.de}}
\par\vspace{1cm}
 Institut f\"ur Theoretische Physik, RWTH Aachen University, 52056 Aachen, Germany
\par\vspace{1cm}
{\bf Abstract}\\
\parbox[t]{\textwidth}
{\small{  
 We compute the  differential and total rate
 of the semileptonic 
 decay  of polarized top-quarks \\
$t\to \ell \nu_\ell + b~{\rm jet} + {\rm jet}$ 
 at next-to-leading order (NLO) in the QCD coupling
  with an off-shell intermediate $W$ boson.
 We present several  normalized distributions,
  in particular those that reflect  
 the $t$-spin analyzing powers of 
the lepton, the b-jet and the $W^+$ boson at LO and NLO QCD.
}}

\end{center}
\vspace*{0.7cm}

PACS number(s): 12.38.Bx,  14.65.Ha
\end{titlepage} 
%
%
\setcounter{footnote}{0}
\renewcommand{\thefootnote}{\arabic{footnote}}
\setcounter{page}{1}

\section{Introduction} 
\label{introduction}

The top quark, the heaviest known fundamental particle, is set apart
from the lighter quarks by the fact that it is so short-lived that
it does not hadronize.
The top quark decays almost exclusively into a $b$ quark and a $W$ boson;
other decay modes have so far not been observed.

Top-quark production and decay has been explored quite in detail
 at the Tevatron and especially at the Large Hadron
Collider (LHC). So far almost all experimental results agree well
 with corresponding Standard Model (SM) predictions.  (For recent overviews,
see \cite{Schilling:2012dx,Deliot:2013gla,EfeYazganfortheCMS:2013zca,Chierici:2013naa}.)
On the theoretical side, significant recent progress includes the
computation of the hadronic $t\bar t$ production cross section
 at next-to-next-to-leading order (NNLO) in the QCD coupling
 $\alpha_s$  \cite{Baernreuther:2012ws,Czakon:2013goa} and the calculation
 of the  differential decay rate of $t\to b\ell\nu_\ell$ at  NNLO 
 in perturbative QCD  \cite{Gao:2012ja,Brucherseifer:2013iv}.

Over the years, top-quark decay has been analyzed in detail within the SM.  As to the
 total decay width $\Gamma_t$, the order $\alpha_s$ QCD corrections\cite{Jezabek:1988iv,Jezabek:1993wk},
 the order $\alpha$ electroweak corrections \cite{Denner:1990ns,Eilam:1991iz}, and the order
 $\alpha_s^2$  QCD corrections \cite{Czarnecki:1998qc,Chet99} were calculated quite some time ago.
 The  fractions of top-quark decay into $W^+$ with helicity
 $\lambda_W=0, \pm1$ are also known to  NNLO QCD \cite{Czarnecki:2010gb}, including 
the order $\alpha$ electroweak corrections \cite{Do:2002ky}. 
 Differential distributions of semileptonic and
 non-leptonic decays of (un)polarized top quarks
 were determined
 to NLO in the gauge couplings
 \cite{Jezabek:1988ja,Czarnecki:1990pe,Brandenburg:2002xr,Fischer:2001gp,Groote:2006kq,Hagiwara:2007sz,Kadeer:2009iw,Ali:2009sm,Kitadono:2012qk},
 and $b$-quark fragmentation was analyzed in \cite{Corcella:2001hz,Cacciari:2002re,Corcella:2009rs,Kniehl:2012mn,Nejad:2013fba}.

In this paper we compute the  differential and total  rate of polarized
top-quarks decaying into $\ell \nu_\ell + b~{\rm jet} + {\rm jet}$ at NLO in the
QCD coupling. The differential rate  is  of interest
 as a building block for predictions of top-quark production and decay
 at NLO QCD, for instance for $t{\bar t} + {\rm jet}$ production
 \cite{Dittmaier:2007wz,Dittmaier:2008uj,Melnikov:2011qx}, for single 
  top-quark + ${\rm jet}$ 
 production at the LHC,  or for $t{\bar t} + {\rm jet}$ production at a future
 $e^+e^-$ linear collider. In fact, this decay mode was already computed to
 NLO QCD by \cite{Melnikov:2011qx}. The results of this paper  on
 $t{\bar t} + {\rm jet}$ production at hadron colliders include also NLO jet radiation 
 in top-quark decay. Distributions for this decay mode were not given separately in \cite{Melnikov:2011qx}.
  Therefore, we believe that it is useful to
 present,  for possible applications to
 other processes, a separate detailed analysis
 of this top-quark decay mode.

The paper is organized as follows. In Sec.~\ref{setup} we describe our computational set-up.
 In Sec.~\ref{sec:obsR} we present our results for the decay rate and for
 a number of distributions for (un)polarized top-quark decays. Sec.~\ref{sec:concl}
 contains a short summary. In the Appendix we list the subtraction terms, for
  the Catani-Seymour subtraction formalism  \cite{Catani:1996vz,Catani:2002hc} with extensions to
    the case of a coloured massive initial state \cite{Campbell:2004ch,Melnikov:2011qx,Melnikov:2011ta},   
   which we use to handle the soft and collinear divergences that appear in the
   real radiation and NLO virtual correction matrix elements.


\section{Set-up of the computation}
\label{setup}

We consider the decay of polarized top quarks to
 leptons, a $b$-jet and an additional jet,
\begin{equation} \label{tlnujet}
  t\rightarrow  W^{*+} + b~{\rm jet} + {\rm jet}
  \rightarrow {\ell}^+  \nu_\ell \, + b~{\rm jet} \, + {\rm jet},
\end{equation}
at NLO QCD, for an off-shell  intermediate $W$ boson.
 The quarks and leptons in the final state are taken to be massless.

At NLO QCD, i.e., at order $\alpha_s^2$, the differential decay rate of \eqref{tlnujet} is determined
 by the amplitudes of the following parton processes:
\begin{equation} \label{tlnujvir}
 t\rightarrow {\ell}^+ \nu_\ell \, +b \, + g \, , 
\end{equation}
 and
\begin{equation} \label{tlnujrea}
 t\rightarrow  {\ell}^+ \nu_\ell \, + b \, + g g \, , \qquad
  t\rightarrow {\ell}^+  \nu_\ell \, + b \, + q{\bar q} \, , \quad q=u,d,s,c,b \, .
\end{equation}
In the case of additional  $b\bar{b}$ 
 production in the real
  radiation process \eqref{tlnujrea}, we take into account 
     only configurations where a $b \bar{b}$ pair
 is unresolved by  the jet algorithm, 
  i.e., we consider in  \eqref{tlnujet}  
  final states where the additional  jet has  zero  $b$-flavour.
To order $\alpha_s^2$, the matrix element of  \eqref{tlnujvir}
is the sum of the Born term $|{\cal M}_B|^2$ and the
 interference $\delta{\cal M}_V$ of the Born and the 1-loop
 amplitude. The calculation of $\delta{\cal M}_V$, using dimensional
 regularisation, is standard. We expressed
  $\delta{\cal M}_V$, using Passarino-Veltman reduction \cite{Passarino:1978jh},
 in terms of scalar one-loop integrals with up to
  four external legs.
 The scalar integrals that  appear in $\delta{\cal M}_V$ 
 are known analytically in $d$ space-time dimensions (cf., for
  instance, \cite{Ellis:2007qk}  and references therein).
  In particular, we extracted  the ultraviolet  (UV) and infrared (IR) poles in
  $\epsilon =(4-d)/2$ that appear in several of these
  scalar integrals  analytically. 
 The  processes \eqref{tlnujrea} are described by tree-level matrix
 elements.

As to renormalization, the top-quark mass is defined in the on-shell
scheme while the QCD coupling $\alpha_s$ is defined in
the $\overline{\rm MS}$ scheme. 

The soft and collinear divergences that appear in the phase space
integrals of the tree-level matrix elements of \eqref{tlnujrea} and in
   $\delta{\cal M}_V$ are handled with the dipole subtraction method
  \cite{Catani:1996vz,Catani:2002hc}  
  and extensions that apply to the decay of a massive quark \cite{Campbell:2004ch,Melnikov:2011qx,Melnikov:2011ta}.
 Details are given in the Appendix.

 We define jets by
  the Durham algorithm \cite{Catani:1991hj}, i.e.,
  we use the  jet metric
\begin{equation} \label{Durham}
Y_{ij}  =  2 \frac{\min{\left(E_i^2,E_j^2\right)}}{m_t^2} \left(1-\cos{\theta_{ij}}\right),
\end{equation}
where $E_i$, $E_j$ are the energies of partons $i,j$ in the final
state, $\theta_{ij}$ is the angle between them, and $m_t$ denotes the
 mass of the top quark. We work in the rest frame of the $t$ quark.
 The jet resolution parameter is denoted by $Y$. An
 unresolved pair of final-state partons  $i,j$ is recombined
 by adding the four-momenta, $k_{(ij)}  =  k_i + k_j$.
The decay  \eqref{tlnujvir}  contributes 
to \eqref{tlnujet}  events with $Y_{bg} > Y$.
 The real radiation processes \eqref{tlnujrea} contribute to
              \eqref{tlnujet}    events
  with one unresolved pair of partons $i,j$, i.e.,  with 
 $Y_{ij} < Y$. The jet distance
 between the  recombined pseudoparticle $(ij)$ and
  the remaining parton $n$ must satisfy $Y_{n(ij)} > Y$.

\section{Results}
\label{sec:obsR}
 As already mentioned, we work in the top-quark  rest frame.
 If we denote the top-spin vector in this frame by ${\bf s}_t$
 (where ${\bf s}_t^2=1$), differential distributions for the
 decay \eqref{tlnujet}
 of a  100 percent polarized ensemble of top quarks
  are of the form
\begin{equation} \label{DGapol}
\frac{d\Gamma}{dO} = A + {\bf B}\cdot{\bf s}_t \, ,
\end{equation}
 where $O$ denotes some observable.
In the fully
 differential case,
 the functions $A$ and ${\bf B}$ (that transform as scalar and vector,
respectively, under spatial rotations) depend on the independent
 kinematical variables of \eqref{tlnujet}, and the vector  {\bf B}
 may be represented as a linear combination of terms proportional
  to the directions of 
 the charged lepton and of the two jets in the final state.
 
Rotational invariance implies that a number of distributions
 hold both for polarized and unpolarized top quarks.
 This includes the distributions that will be presented in
 Sec.~\ref{susec:unp}.

In Sec.~\ref{susec:pol} we consider distributions that
 are relevant for the decay of polarized top quarks, namely those that reflect
 the top-spin analyzing power of the charged lepton, the the b-jet,
   and the $W$ boson.

For the numerial results given below, we use $m_t= 173.5$ GeV,
 $m_W =80.39$ GeV and $\Gamma_W = 2.08$ GeV. The QCD coupling for
 5-flavour QCD is taken to be $\alpha_s(m_Z)=0.118$. Its evolution
 to $\mu=m_t$ and conversion to the 6-flavour  $\overline{\rm MS}$
 coupling results in  $\alpha_s(m_t)=0.108$. Moreover, we use 
 $\alpha(m_t)=7.9\times 10^{-3}$ and $\sin^2\theta_W=0.231$ which
 yields the weak coupling  $g_W^2=0.429$. The normalized
  distributions given below do not depend on $g_W^2$ because we work
 to lowest order in  $g_W^2$.

\subsection{Distributions for (un)polarized top-quark decay}
\label{susec:unp}

\begin{figure}
\centering
\includegraphics[width=12cm,height=12cm]{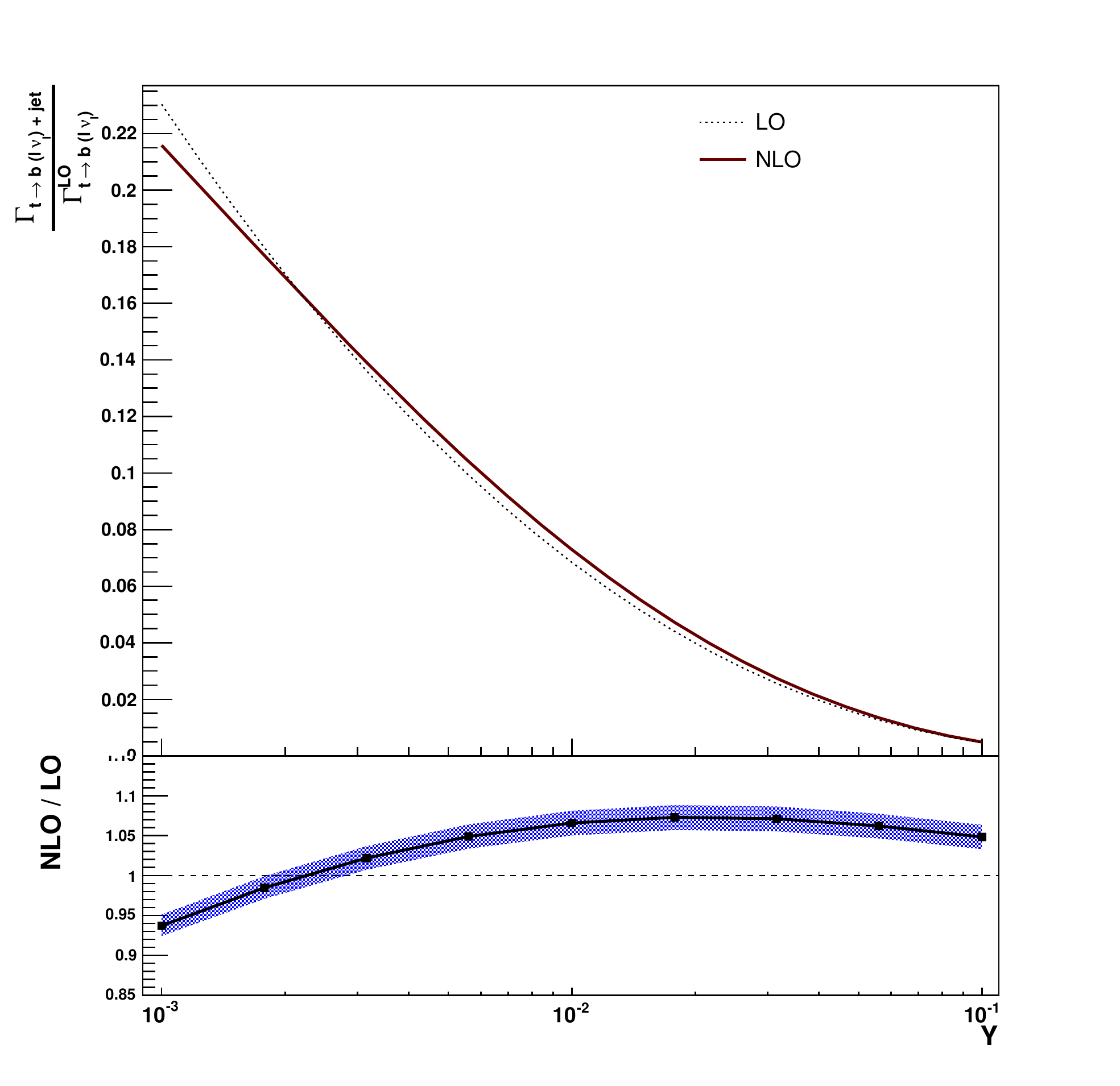}
\caption{Upper pane:
 decay rate  $\Gamma_{t \rightarrow b \, \bar{l}\nu_l
 \, +  \, {\rm jet}}$ (LO and NLO) normalized 
to $\Gamma_{t \rightarrow b \, \bar{l}\nu_l}$ (LO) 
 as a function of the jet resolution parameter $Y$ for
 $\mu=m_t$. Lower pane: ratio of
 $\Gamma_{t \rightarrow b \, \bar{l}\nu_l
 \, +  \, {\rm jet}}^{\rm NLO}/\Gamma_{t \rightarrow b \, \bar{l}\nu_l
 \, +  \, {\rm jet}}^{\rm LO}$  as a function of $Y$. The solid
 line corresponds to $\mu=m_t$, the shaded band results from scale
 variations between $m_t/2$ and $2 m_t$.}
\label{fig:NLO_DECAY_RATE}
\end{figure}

First, we compute the decay rate of  \eqref{tlnujet} as a function
 of the jet resolution parameter $Y$. In Fig.~\ref{fig:NLO_DECAY_RATE}
 the ratio $\Gamma_{t \rightarrow b \, \bar{l}\nu_l \, +  \, {\rm
     jet}}$
 is shown at LO and NLO QCD, normalized to the leading order
 rate $\Gamma_{t \rightarrow b \, \bar{l}\nu_l} = 1.8698 \cdot 10^{-3} \, m_t$,
 for a renormalization scale $\mu=m_t$.  It is clear that
 this ratio increases for decreasing $Y$. 
 In the lower pane of this figure, the `K factor'
$\Gamma_{t \rightarrow b \, \bar{l}\nu_l
 \, +  \, {\rm jet}}^{\rm NLO}/\Gamma_{t \rightarrow b \, \bar{l}\nu_l
 \, +  \, {\rm jet}}^{\rm LO}$ is displayed. One sees that
 in a large range of $Y$, the QCD corrections are positive and at most
 of order $8\%$, while for $Y$ below $\sim 2.5\cdot 10^{-3}$ they become negative.

In the remainder of this section we compute normalized
 decay distributions, both at LO and NLO QCD for 
 two values of the jet resolution parameter,
 $Y=0.01$ and $Y=0.001$. The NLO decay distributions, which are
 normalized to the NLO decay rate $\Gamma_{t \rightarrow b \, \bar{l}\nu_l
 \, +  \, {\rm jet}}^{\rm NLO}$, are expanded in powers of $\alpha_s$.
 Taking out a factor of $\alpha_s$ both from the LO and NLO
  (differential) rate, we have
\begin{eqnarray} \label{exprat}
\frac{{d\Gamma}^{\rm NLO}}{{\Gamma}^{\rm NLO}} & = & \frac{d\Gamma_{0}
  + \alpha_s d\Gamma_{1} + {\cal O}(\alpha_s^2)}{\Gamma_{0} + \alpha_s
  \Gamma_{1} + {\cal O}(\alpha_s^2)}
  =
  \frac{d\Gamma_{0}}{\Gamma_{0}}\left(1-\alpha_s\frac{\Gamma_{1}}{\Gamma_{0}}\right) 
 + \alpha_s\frac{d\Gamma_1}{\Gamma_0} + {\cal O}(\alpha_s^2). \nonumber
\end{eqnarray}
 
In the following we rescale all dimensionful variables with
$m_t$. That is, in the following, 
the energies $E_W$, $E_l$, $E_b$, and $E_2$  of the $W$ boson, the charged
lepton,  $b$-jet, and the second jet with zero $b$-flavor, respectively, 
  and the $W$ and $\ell b$-jet invariant masses $M_W$, $M_{lb}$ denote
dimensionless variables.

\begin{figure}
\centering
\includegraphics[angle=0,width=1.0\textwidth]{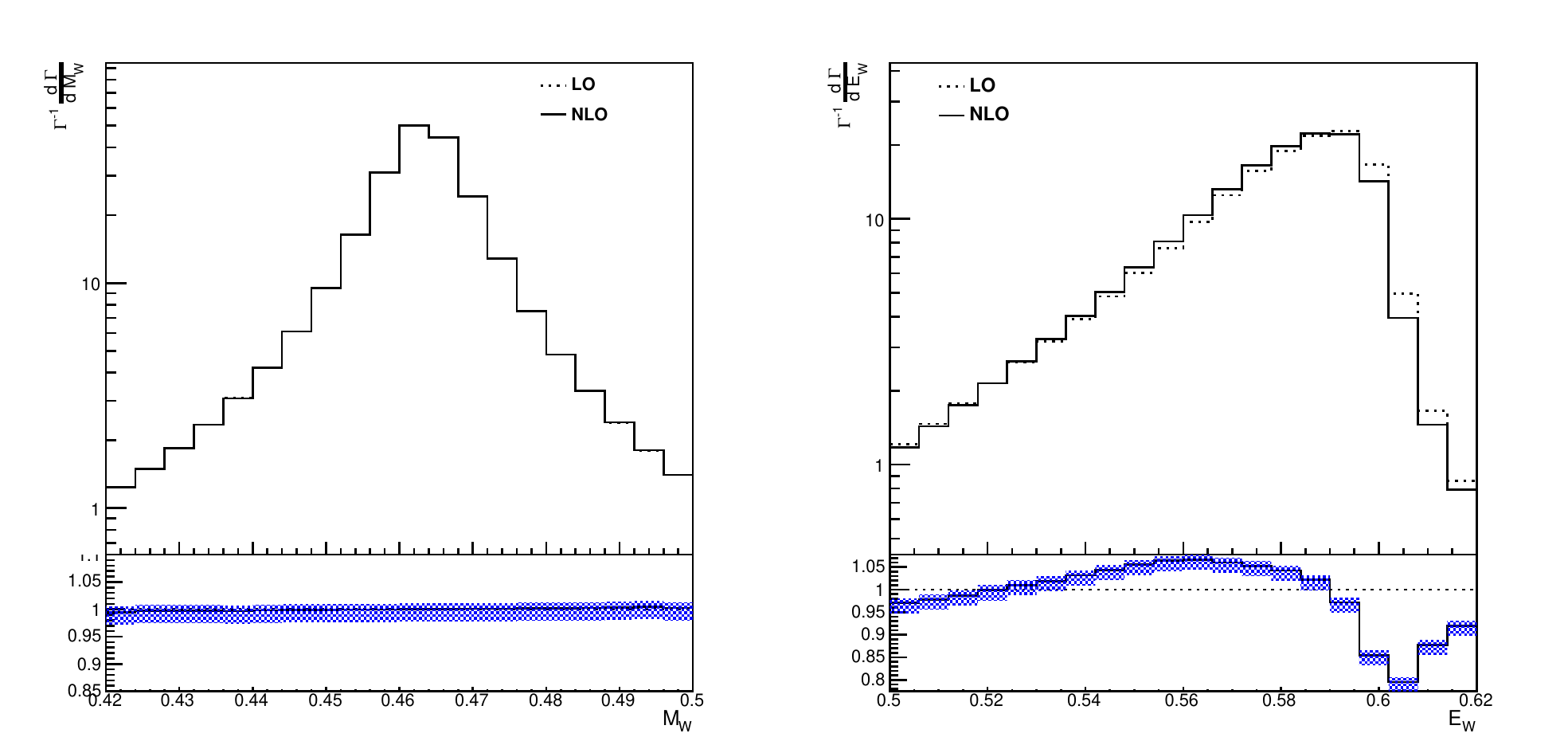}
\caption{Upper panes:
 normalized distribution of 
 the invariant mass $M_W$ of the
 $W$ boson (left) and of the $W$ energy $E_W$ (right)
 for $Y=0.01$ and $\mu=m_t$.
  Lower panes:
 ratio of the NLO and corresponding LO distribution. The shaded band results from scale
 variations between $m_t/2$ and $2 m_t$.}
\label{fig:NLO_DIFF_001_1}
\end{figure}

\begin{figure}
\centering
\includegraphics[angle=0,width=1.0\textwidth]{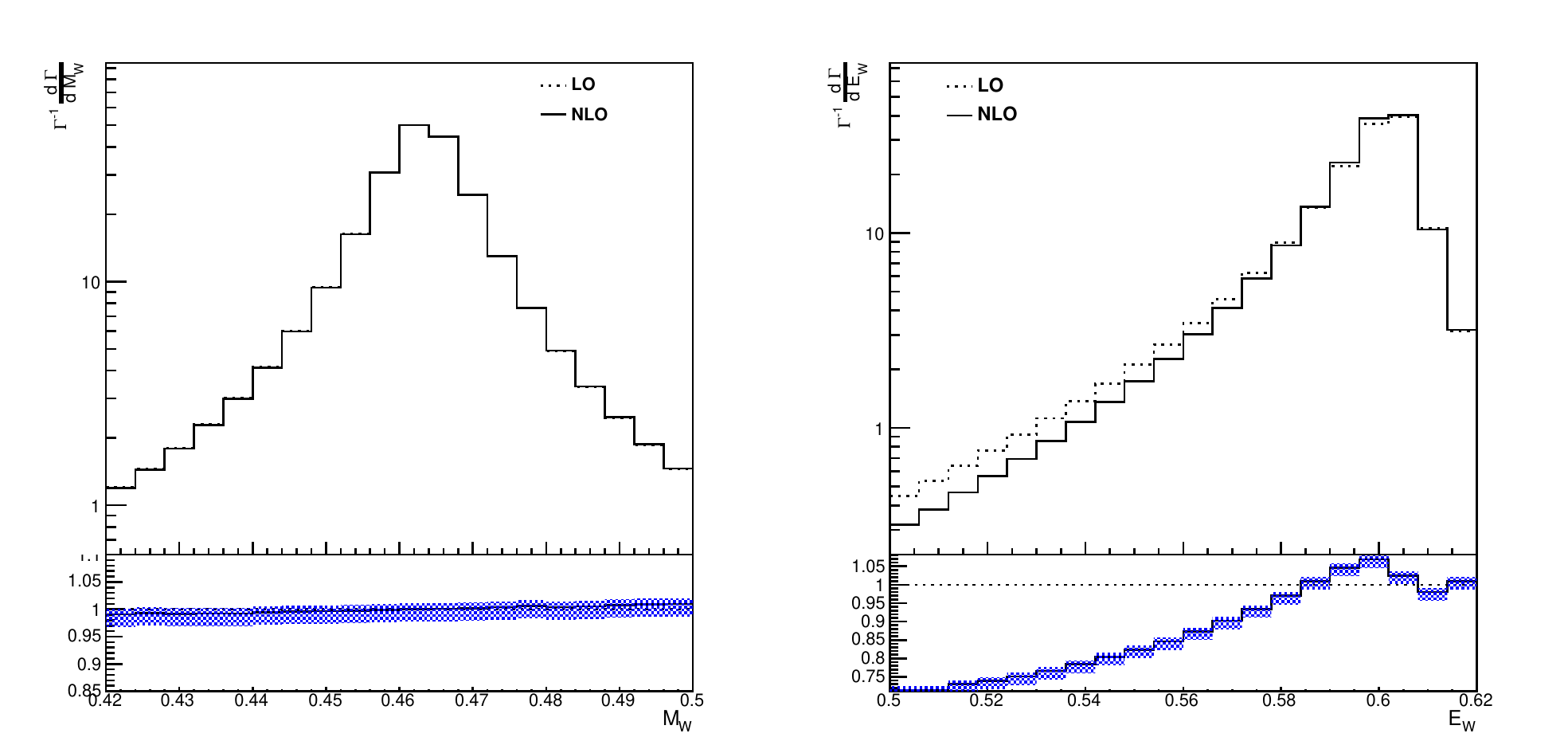}
\caption{Same as Fig.~\ref{fig:NLO_DIFF_001_1}, but
 for a jet resolution parameter $Y=0.001$.}
\label{fig:NLO_DIFF_0001_1}
\end{figure}

 The invariant mass distribution and
the energy distribution of the off-shell $W$ boson are displayed 
  in Fig.~\ref{fig:NLO_DIFF_001_1} and~\ref{fig:NLO_DIFF_0001_1}
  for $Y=0.01$ and $Y=0.001$, respectively. 
 The QCD corrections to the invariant mass of the $W$ boson are very
 small. The distribution of the $W$ energy\footnote{Here, we tacitly assume
   that the neutrino energy and momentum can be reconstructed in an experiment, 
   which is usually possible only with ambiguities.}  $E_W=E_l+E_\nu$
 may be compared with the case of the lowest-order on-shell
 decay $t\to b W$ where the  (dimensionless)  $W$ energy is
 fixed, ${\bar E}_W =\sqrt{m_W^2+{\bf k}_W^2}/m_t =0.61.$
 In the case of additional jet radiation and allowing the $W$ boson to
 be off-shell, one expects therefore that the maximum of the
   distribution of $E_W$ is below ${\bar E}_W$, but approaches this
   value if the  jet cut $Y$ is decreased. The distributions on the right
   sides of   Figs.~\ref{fig:NLO_DIFF_001_1} and~\ref{fig:NLO_DIFF_0001_1} show
  this behaviour. The QCD corrections are small at and in
  the near vicinity of the maximum of the distribution, whereas they
  can become rather large if the $W$ boson is significantly off-shell.

\begin{figure}
\centering
\includegraphics[angle=0,width=1.0\textwidth]{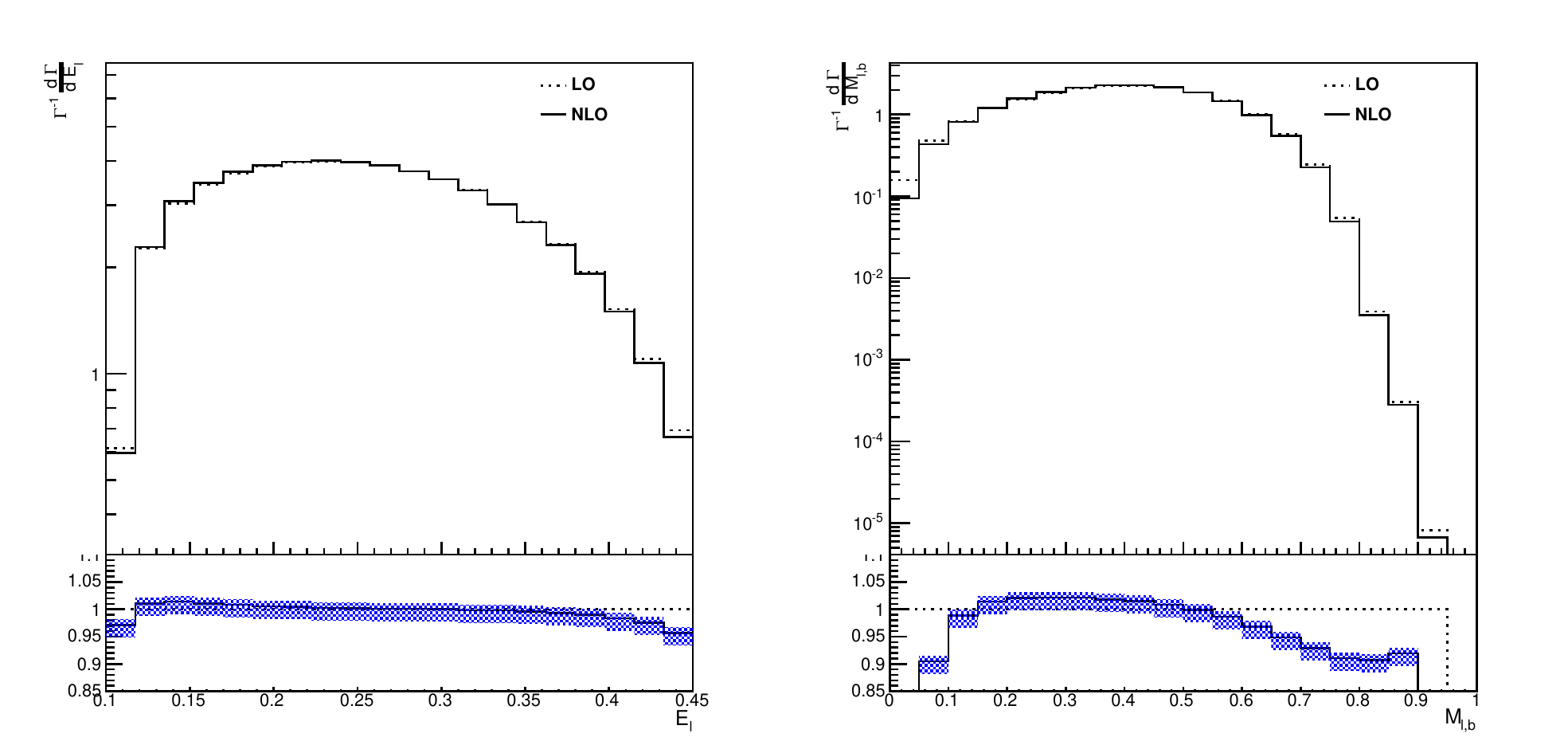}
\caption{Upper panes: 
 normalized
 distribution of the  energy $E_l$ of the 
 charged lepton  (left) and 
 of the
 invariant mass $M_{lb}$ 
 of of the $b$ jet and the charged lepton (right)
 for  $Y=0.01$ and $\mu=m_t$.
  Lower panes:   
 ratio of the NLO and corresponding LO distribution.  The shaded band results from scale
 variations between $m_t/2$ and $2 m_t$.  }
\label{fig:NLO_DIFF_001_2}
\end{figure}

\begin{figure}
\centering
\includegraphics[angle=0,width=1.0\textwidth]{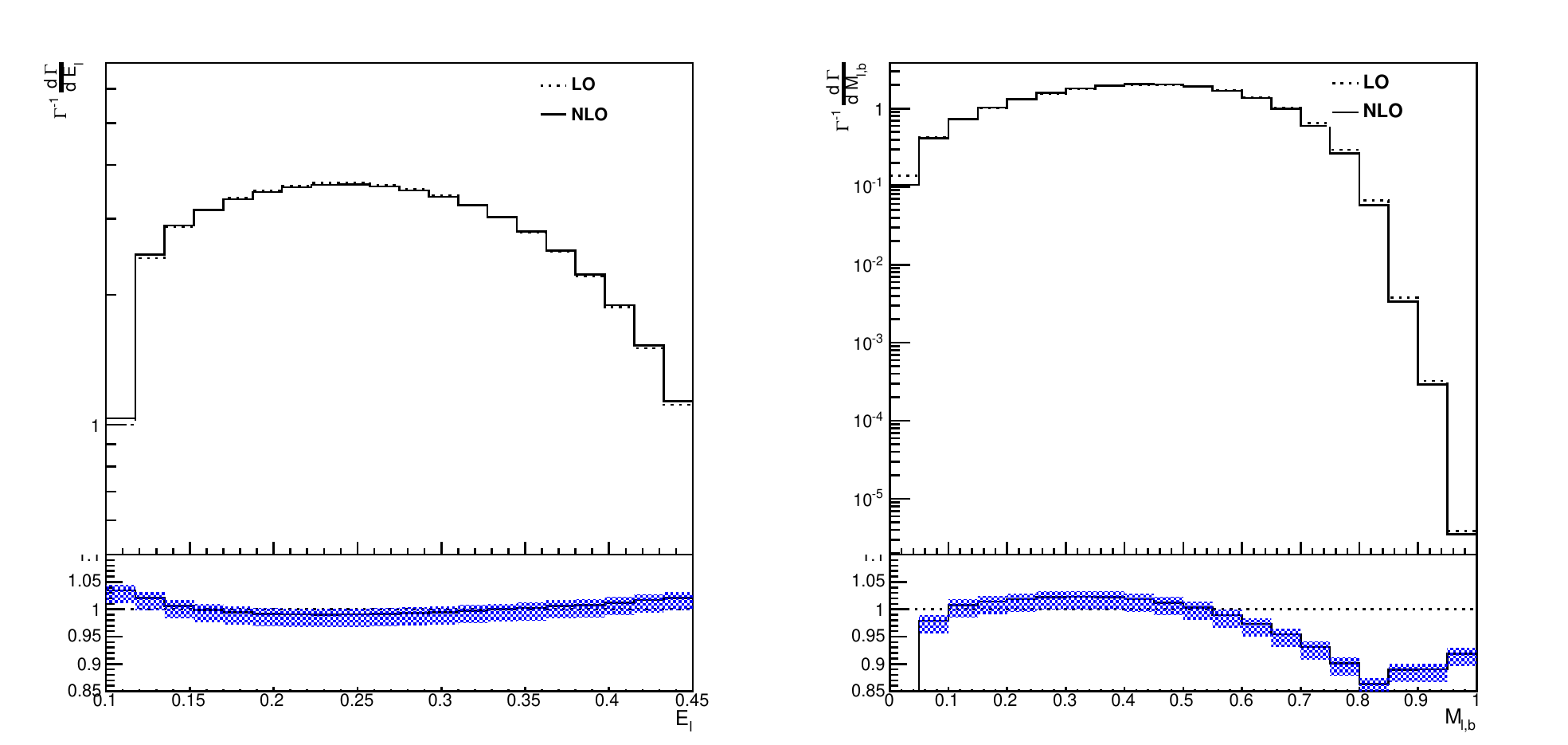}
\caption{Same as Fig.~\ref{fig:NLO_DIFF_001_2}, but for $Y=0.001$.}
\label{fig:NLO_DIFF_0001_2}
\end{figure}

The left sides of Figs.~\ref{fig:NLO_DIFF_001_2} and~\ref{fig:NLO_DIFF_0001_2}
show the  distribution of the  energy $E_l$ of the 
 charged lepton. For decreasing jet cut the distribution moves towards
 the  lepton-energy distribution of
  the inclusive semileptonic decay  which, at tree level and for
 a massless $b$ quark, has  its maximum
  at $E_l=0.25$.  

The right sides of Figs.~\ref{fig:NLO_DIFF_001_2} and~\ref{fig:NLO_DIFF_0001_2}
 display the  distribution of the invariant mass $M_{lb}$ of the
  lepton and the $b$ jet\footnote{The distribution of
   $M_{lb}$ in inclusive hadronic $t\bar t$ production and decay was
   first analyzed at NLO QCD in \cite{Melnikov:2009dn} and was proposed as a tool to measure the top-quark
   mass. Cf. also   \cite{Denner:2012yc}.}. In the case of the LO decay
   $t\to b\ell\nu_\ell$ and an on-shell intermediate $W$ boson,
    $M_{lb}$ has a sharp upper bound, which, in terms of our
    dimensionless variables, is given by $M_{lb}^{\rm max}=\sqrt{1-m_W^2/m_t^2}$. 
    In the case  of \eqref{tlnujvir}, \eqref{tlnujrea},
  where gluons   or $q\bar q$ are radiated, 
 the invariant mass  $M_{lb}$ cannot exceed the LO kinematic boundary,
 as long as the $W$ boson is kept on-shell. The distance between the
 maximum of the  $M_{lb}$ distribution and $M_{lb}^{\rm max}$ is
 expected to decrease with decreasing jet cut $Y.$
 An off-shell $W$ boson  leads to a tail of the $M_{lb}$ distribution beyond 
 $M_{lb}^{\rm max}$. All of these features arise in the results shown 
 on the right sides of Figs.~\ref{fig:NLO_DIFF_001_2}
 and~\ref{fig:NLO_DIFF_0001_2}. In the vicinity of $M_{lb}^{\rm max}=0.89$
 the QCD corrections are about $-10\%$. \\

\begin{figure}
\centering
\includegraphics[angle=0,width=1.0\textwidth]{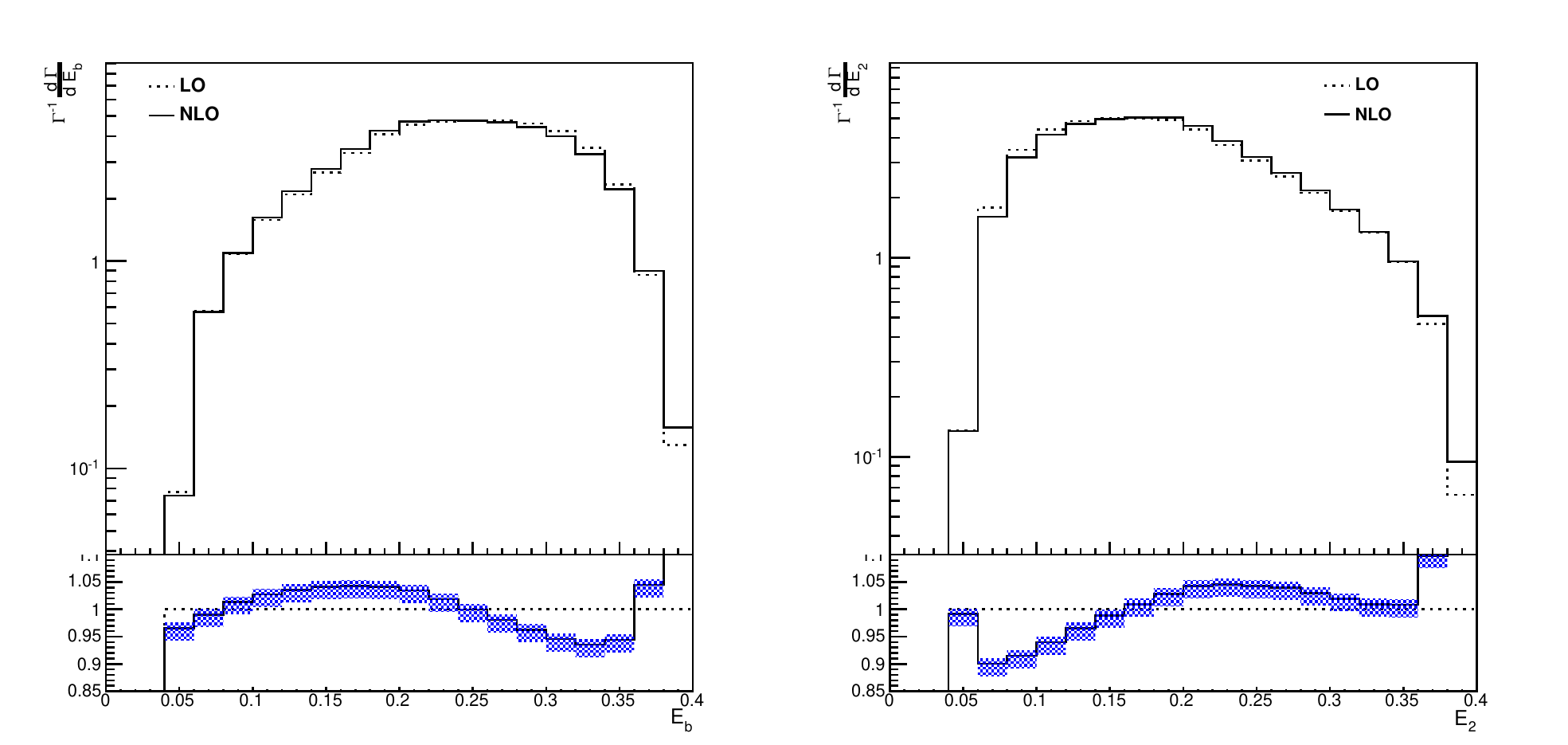}
\caption{Upper panes: normalized 
 distribution of b-jet energy $E_b$ (left) and of the energy
 $E_2$ of the second jet (right) for $Y=0.01$ and $\mu=m_t$.
 Lower panes: ratio of the NLO and corresponding LO distribution.  The shaded band results from scale
 variations between $m_t/2$ and $2 m_t$. }
\label{fig:NLO_DIFF_001_3}
\end{figure}

\begin{figure}
\centering
\includegraphics[angle=0,width=1.0\textwidth]{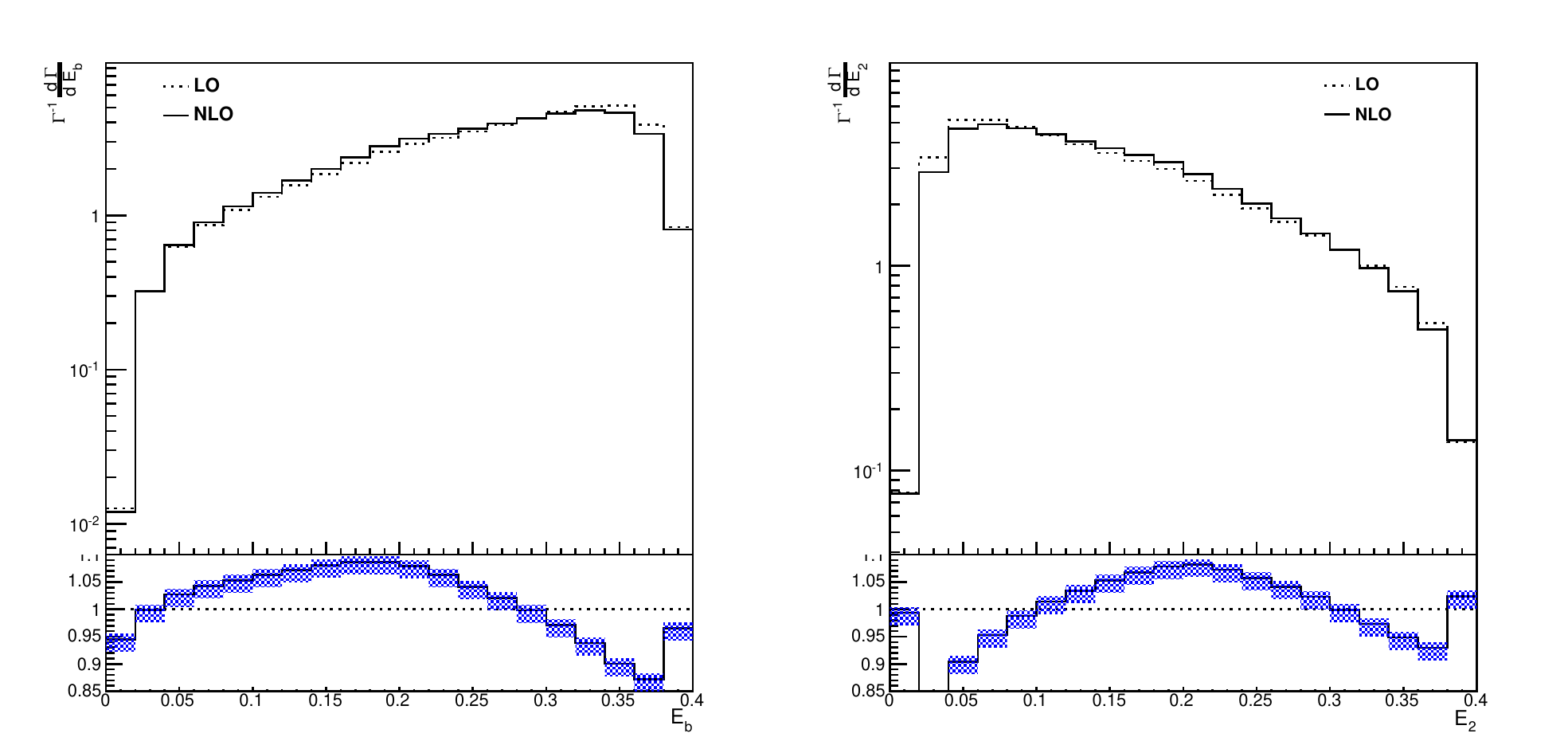}
\caption{Same as Fig.~\ref{fig:NLO_DIFF_001_3}, but for $Y=0.001$.}
\label{fig:NLO_DIFF_0001_3}
\end{figure}

The distribution of the b-jet energy $E_b$  and of the energy
 $E_2$ of the second jet is displayed in Figs.~\ref{fig:NLO_DIFF_001_3}
and~\ref{fig:NLO_DIFF_0001_3}. In the case of  the LO decay $t\to b
W$, the energy of the massless $b$ quark is fixed to be ${\bar E}_b
=(1-m_W^2/m_t^2)/2 =0.39$.  Radiation off the $t$ and $b$ leads to an
upper bound on $E_b$ that is below  ${\bar E}_b$ for $Y>0$. An
off-shell $W$ boson can, however, lead to some events with  $E_b$
above this value. The average energy $E_2$ of the second jet is
smaller than  that of the $b$ jet. These features are exhibited by the
 results shown in  Figs.~\ref{fig:NLO_DIFF_001_3}
and~\ref{fig:NLO_DIFF_0001_3}. Near the kinematic edges the QCD
corrections can become $\sim 10\%$.

\begin{figure}
\centering
\includegraphics[angle=0,width=1.0\textwidth]{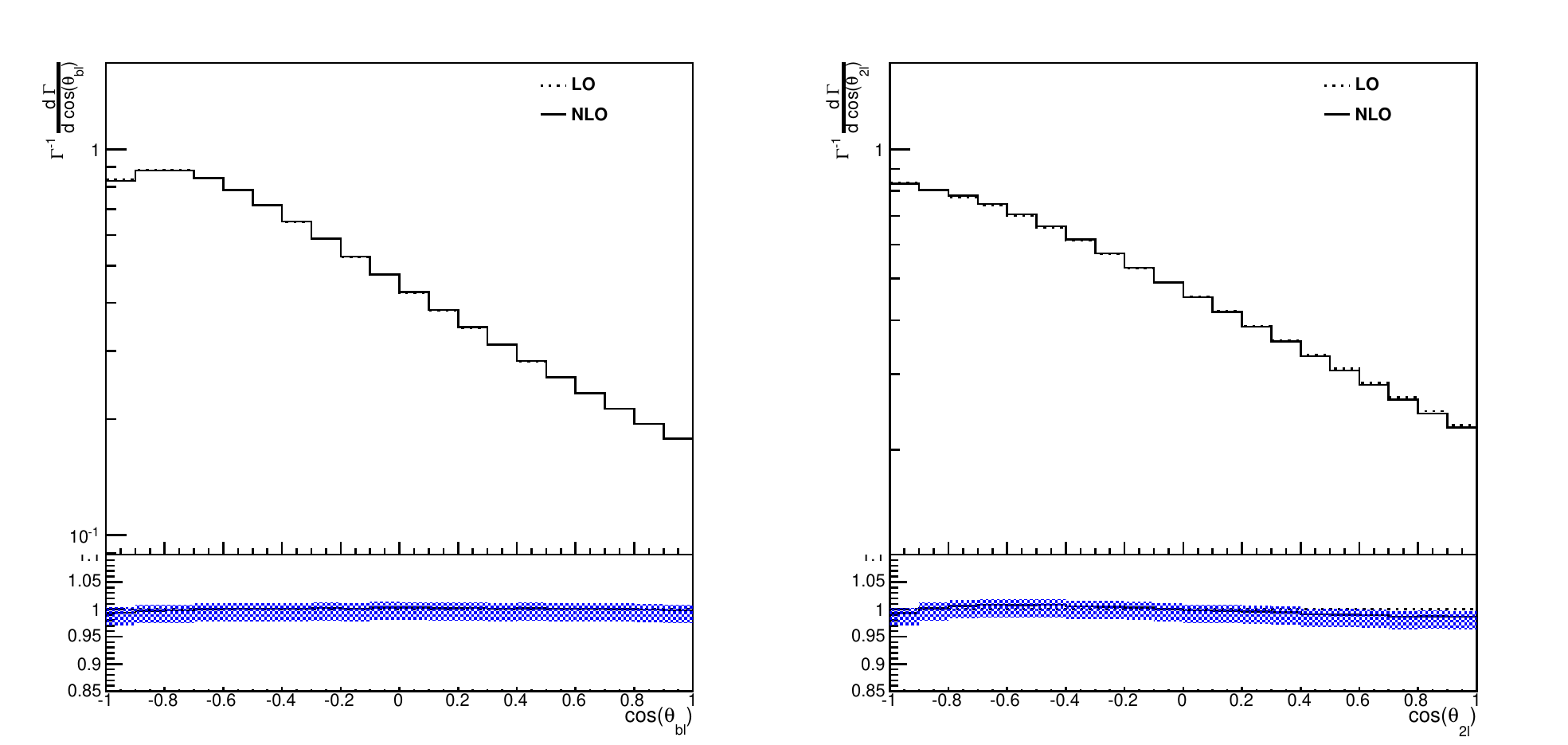}
\caption{Upper
 panes: 
 normalized distribution 
 of  $\cos{\theta_{bl}}$, where $\theta_{bl}$ is the
  angle between the directions of flight of the charged lepton
  and the b-jet in the $t$ rest frame (left), and of  $\cos{\theta_{2l}}$, where $\theta_{2l}$ is the
  angle between    $\ell^+$ and  the second jet.
The jet resolution parameter is chosen to be $Y=0.01$ and $\mu=m_t$.
  Lower panes: ratio of the NLO and corresponding LO distribution.  The shaded band results from scale
 variations between $m_t/2$ and $2 m_t$.}
\label{fig:NLO_DIFF_001_4}
\end{figure}

\begin{figure}
\centering
\includegraphics[angle=0,width=1.0\textwidth]{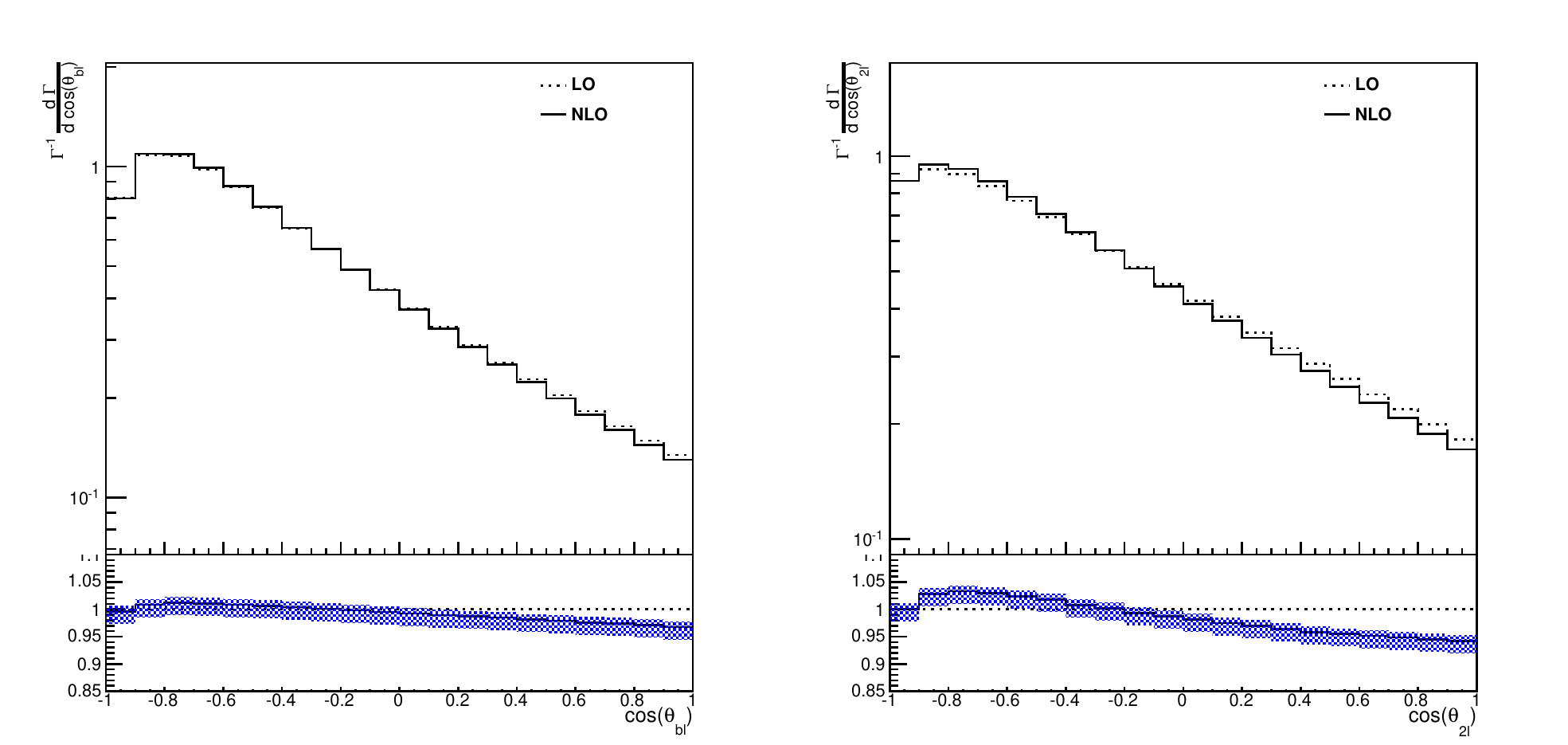}
\caption{Same as Fig.~\ref{fig:NLO_DIFF_001_4}, but for $Y=0.001$.}
\label{fig:NLO_DIFF_0001_4}
\end{figure}

 Figs.~\ref{fig:NLO_DIFF_001_4} and~\ref{fig:NLO_DIFF_0001_4}
 show the distribution 
 of  $\cos{\theta_{bl}}$, where $\theta_{bl}$ is the
  angle between the directions of flight of the charged lepton
  and the b-jet in the $t$ rest frame, and of  $\cos{\theta_{2l}}$, where $\theta_{2l}$ is the 
  angle between  $\ell^+$  and  the second jet.
 The distributions of $\cos{\theta_{bl}}$ are qualitatively similar to
 the corresponding distributions in the case of inclusive semileptonic
 top-decay; for most of the events the charged lepton and the $b$ jet
 are almost back-to-back. As expected, the distribution of $\cos{\theta_{2l}}$ is
  falling less steeply towards smaller angles
 $\theta_{2l}$. The QCD corrections are markedly below $5\%$  in most of the kinematic
 range.

\begin{figure}
\centering
\includegraphics[angle=0,width=1.0\textwidth]{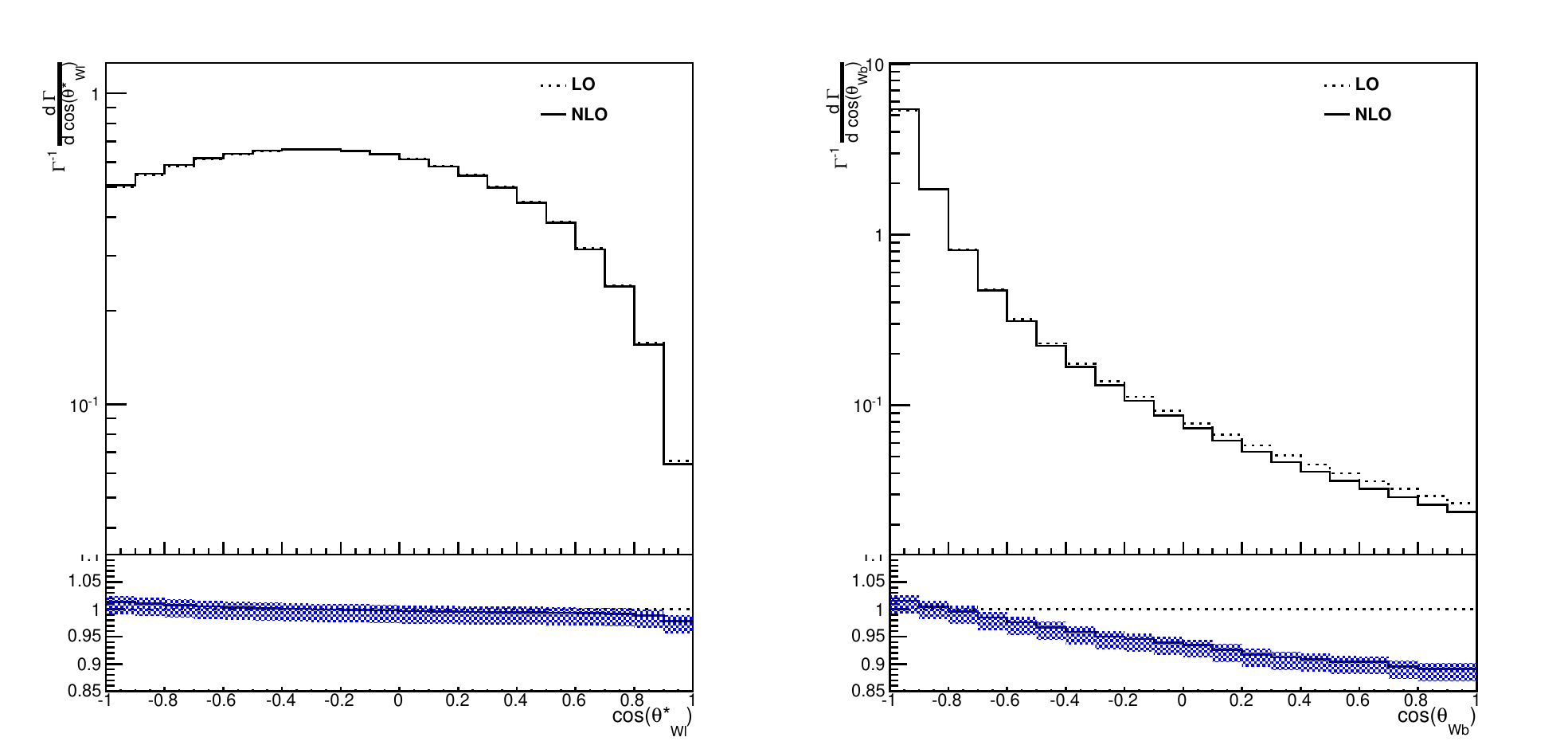}
\caption{Upper panes:
 normalized   distribution  of
 $\cos\theta^*_{Wl}$, where $\theta^*_{Wl}$ is the angle between the $W^+$
 direction in the t rest frame and the lepton direction in the
 $W^+$ rest frame (left). 
       The right plot shows the normalized  distribution
       of $\cos{\theta_{Wb}}$, where $\theta_{Wb}$ is the angle
       between the  $W$ and the b-jet directions in the $t$ rest frame.
  The jet resolution parameter  is  $Y=0.01$  and  $\mu=m_t$.
 Lower panes: ratio of the NLO and corresponding LO distribution.  The shaded band results from scale
 variations between $m_t/2$ and $2 m_t$.}
\label{fig:NLO_DIFF_001_5}
\end{figure}

\begin{figure}
\centering

\includegraphics[angle=0,width=1.0\textwidth]{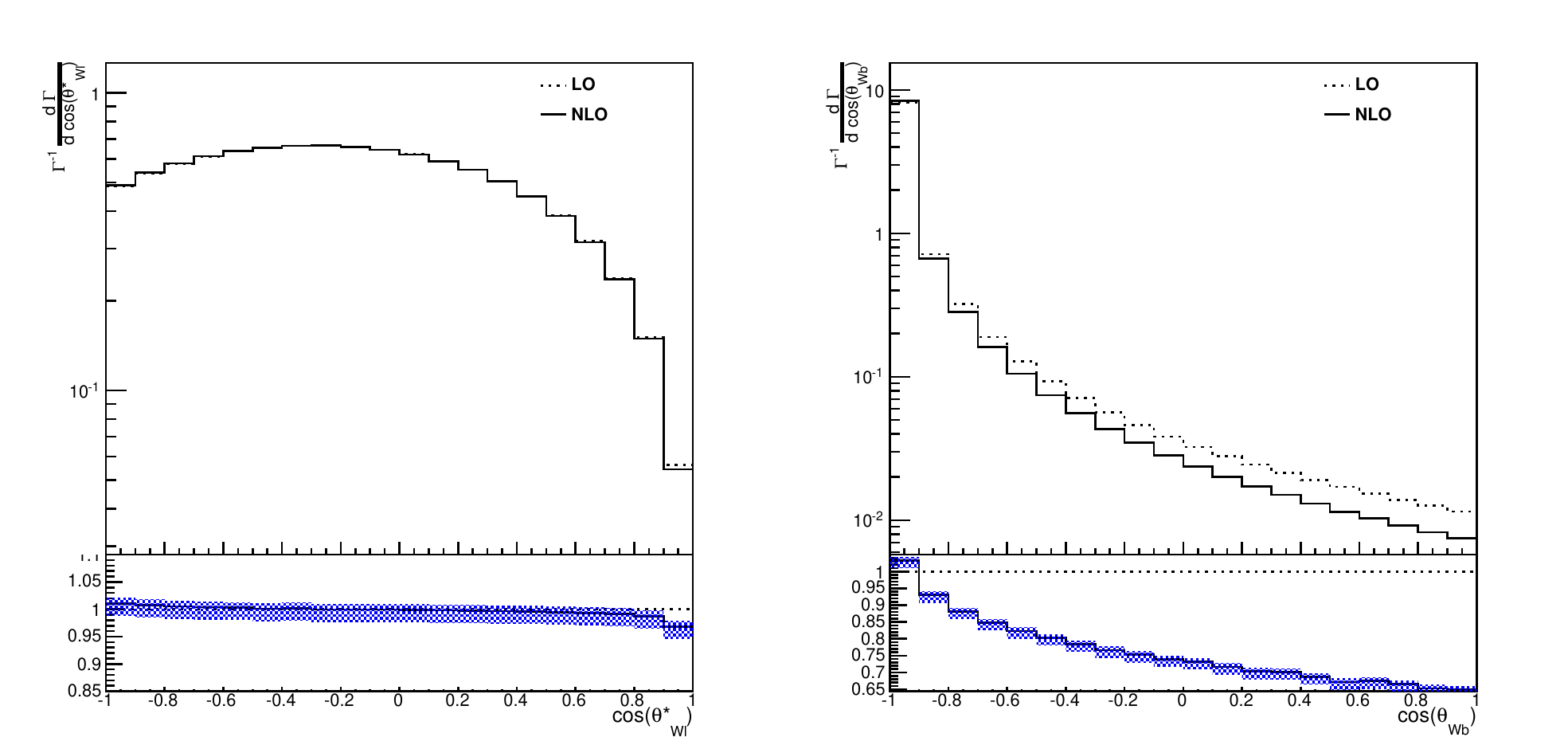}
\caption{Same as Fig.~\ref{fig:NLO_DIFF_001_5}, but for $Y=0.001$.}
\label{fig:NLO_DIFF_0001_5}
\end{figure}

The distribution  of
 $\cos\theta^*_{Wl}$, where
 $\theta^*_{Wl}$ is the angle between the $W^+$
 direction in the t rest frame and the lepton direction in the
 $W^+$ rest frame,  
  is presented in the  plots on the left side of 
     Figs.~\ref{fig:NLO_DIFF_001_5}  
  and~\ref{fig:NLO_DIFF_0001_5}.
 This distribution has been used ever since at the
 Tevatron and the LHC  for measuring the $W$-boson helicity fractions 
 in  inclusive semileptonic top-decay. With $x=\cos\theta^*_{Wl}$
 the one-dimensional distribution has the well-known form
 \[ \Gamma^{-1}\frac{d\Gamma}{dx} =
 \frac{3}{4}F_L\left(1-x^2\right) + \frac{3}{8}F_-\left(1-x\right)^2 
  + \frac{3}{8}F_+\left(1+x\right)^2 \, , \]
with $F_L + F_- + F_+= 1$.
 For events with an additional jet, one expects that for small jet cut
 $Y$ the corresponding distribution tends towards the inclusive
 one.  
 Performing a  fit to the $\cos\theta^*_{Wl}$ distributions of Figs.~\ref{fig:NLO_DIFF_001_5}  
  and~\ref{fig:NLO_DIFF_0001_5} (where we take into
 account that our NLO distributions are not exactly normalized to one, due
 to the expansion \eqref{exprat}), we obtain
 $F_L^{\rm NLO}= 0.668$ and  $F_-^{\rm NLO}=0.321$ for  $Y=0.01$,
 and  $F_L^{\rm NLO}= 0.689$ and  $F_-^{\rm NLO}=0.308$ for  $Y=0.001$.
 The  size of the QCD corrections is $\lesssim 1\%$. For  $Y=0.001$ the helicity
 fractions agree very well with the corresponding inclusive ones
 at NLO QCD (cf., for instance, \cite{Czarnecki:2010gb}) and are in
 agreement with recent results from ATLAS and CMS \cite{Kroening}.

 The plots on the right sides of Figs.~\ref{fig:NLO_DIFF_001_5}  
  and~\ref{fig:NLO_DIFF_0001_5}
  show the distribution
       of $\cos{\theta_{Wb}}$, where $\theta_{Wb}$ is the angle
       between the  $W$ and the b-jet directions in the $t$ rest
       frame. As in the inclusive case this distribution peaks when
       the $W$ boson and the $b$ jet are back-to-back.

\clearpage

\clearpage

\clearpage

\clearpage

\clearpage

\subsection{Top-spin analyzing power}
\label{susec:pol}
Finally we consider, for the decay  \eqref{tlnujet} of a $100 \%$  polarized
top-quark ensemble,  the angular correlation  of the
 top-spin vector ${\bf
  s}_t$ and the direction of flight of 
 a final-state particle or jet $f$ 
 in the top rest frame, where $f=\ell^+,~b~{\rm jet},~W^+$.
 The corresponding normalized distribution 
  has the a priori form
\begin{equation}\label{topspinpow}
\frac{1}{\Gamma}\frac{d\Gamma}{d\cos{\theta_{f}}}  =  \frac{1}{2}
\left( 1 + \kappa_f \cos{\theta_{f}} \right) \, ,
\end{equation}
where $\theta_{f} =\angle({\bf s}_t, {\bf\hat k}_f)$.
The coefficient $\kappa_f$ is the top-spin
 analyzing power of $f$ and  measures the degree of correlation.
  CP invariance implies\footnote{The
   effect of the non-zero Kobayashi-Maskawa phase, which would
  show up only if higher order weak corrections are taken into account,
     is completely
   negligible in these decays.} that the corresponding angular distributions
   for top antiquarks are given by
\begin{equation}\label{atopspinpo}
\frac{1}{\bar\Gamma}\frac{d\bar\Gamma}{d\cos{\theta_{\bar f}}}  =  \frac{1}{2}
\left( 1  - \kappa_f \cos{\theta_{\bar f}} \right) \, .
\end{equation}
The values of $\kappa_f$ can be extracted from the slope of the
  distributions \eqref{topspinpow}
  or from
 \begin{equation}\label{kapav}
\langle \cos\theta_f \rangle =
  \frac{1}{\Gamma}\int_0^\pi d\cos{\theta_{f}} \,
 \left( \cos{\theta_{f}} \, \frac{d\Gamma}{d\cos{\theta_{f}} } \right)  =  \frac{\kappa_f}{3} \, .
\end{equation}

The results for $\kappa_f$ at LO and NLO QCD
  are listed in Table \ref{tab:spin_correlation_factors} for
  two values of the jet resolution parameter $Y$.

\vspace{0.5cm}
\begin{table}[ht]
\begin{center}
\caption{Top-spin analyzing powers extracted from the normalized
  distributions (\ref{topspinpow})
 for $\mu=m_t$.  The uncertainties due
  to scale variations between $m_t/2$ and $2 m_t$ are below 1\%.}
\vspace{2mm}
\begin{tabular}{|l|l|l|} \hline
  & \multicolumn{1}{c|}{Y=0.01} & \multicolumn{1}{c|}{Y=0.001}  \\ \hline
  $\kappa_\ell^{\rm LO}  $  &   0.981   &  0.993 \\ 
  $\kappa_\ell^{\rm NLO} $  &   0.983   &  0.996  \\ \hline
  $\kappa_W^{\rm LO}    $  &   0.359   &  0.387  \\ 
  $\kappa_W^{\rm NLO}   $  &   0.351   &  0.381  \\ \hline
  $\kappa_b^{\rm LO}    $  &  -0.326   &  -0.368 \\ 
  $\kappa_b^{\rm NLO}   $  &  -0.319   &  -0.364 \\ \hline
\end{tabular}
\label{tab:spin_correlation_factors}
\end{center}
\end{table}

One may compare these $t$-spin analyzing powers with
 the corresponding ones of the dominant
 semileptonic decay modes $t\to b \ell^+ \nu_\ell.$
In the latter case one has $\kappa_\ell^{\rm NLO}=0.999$
\cite{Czarnecki:1990pe} and
  $\kappa_b^{\rm NLO} = -0.39$
   \cite{Brandenburg:2002xr}.
   Moreover, in this inclusive case,
 $\kappa_b^{\rm NLO} = -\kappa_W^{\rm NLO}$.
 The charged lepton is the best top-spin analyzer in the
semileptonic decays both without and with an additional jet. This 
  is due to the V-A structure of the charged weak current and angular
 momentum conservation. 
 If an additional jet is produced
  in top quark decay, $\kappa_b = -  \kappa_W$ no longer holds, of
  course, cf. Table~\ref{tab:spin_correlation_factors}.
 In semileptonic $t$ decays both without and with an additional jet
 the  $t$-spin analyzing power of the $W$ boson is weaker than that
  of its daughter lepton $\ell^+$. This is due to the known fact that 
  for $t\to \ell^+ \nu_\ell b~(+{\rm jet})$, the amplitudes that correspond
  to the different polarization states of the intermediate $W$ boson
 interfere constructively (destructively) when $\ell^+$ is emitted in
 (opposite to) the direction of the top spin.

\section{Summary}
\label{sec:concl}

We have computed  the  differential and total rate
 of the semileptonic  decay  of polarized top-quarks
$t\to \ell \nu_\ell + b~{\rm jet} + {\rm jet}$ 
 at next-to-leading order QCD. We have defined the jets by the Durham
 algorithm, and we have presented a number of distributions for two
 different values of the jet resolution parameter.
 The QCD corrections to the leading-order distributions are 
 $\lesssim 5\%$ in most of the kinematic range. Near kinematic edges or
 significantly off the $W$ resonance, the corrections can become
 $\sim 10\%$. Our results should be useful as a building block for
 future analyses of top-quark production and decay in hadron and 
 in $e^+e^-$ collisions.
 

\subsubsection*{Acknowledgements}

  The work of  W.B. was supported by BMBF and that of
  C.M. by Deutsche Forschungsgemeinschaft through Graduiertenkolleg GRK
  1675.

\section*{Appendix}

In this appendix we collect, for the convenience of the reader, the
 un-integrated and integrated subtraction terms that we used to
 handle the soft and collinear divergences which appear
 in the phase space integrals of the real radiation matrix
  elements of the processes \eqref{tlnujrea}
 and in the 1-loop corrections to
 \eqref{tlnujvir}. We use the dipole subtraction method
 \cite{Catani:1996vz}
 and its extension to the decay of a massive quark, worked out
  in  \cite{Campbell:2004ch,Melnikov:2011ta,Melnikov:2011qx}.

In this framework, the decay rate of  \eqref{tlnujet} is given, as a function
 of the jet resolution parameter $Y$, at NLO QCD by
 $\Gamma^{\rm NLO}(Y)= \Gamma^{\rm B}(Y) +\delta\Gamma(Y)$, where
\begin{align}
 \delta \Gamma(Y)  & =  \int d\phi_4 \, \left( \delta{\cal
     M}_V^{(d)}F_4(Y) + F_4(Y)\int d\phi^{\rm (dip.)} \, \delta{\cal
     M}_{\rm CT}^{(d)} \right)_{d=4}  \nonumber \\
& + \int d\phi_5 \, \left(({\cal M}^*_R {\cal M}_R)^{(4)} F_5(Y) -
  \tilde{F}_4(Y) \delta{\cal M}_{\rm CT}^{(4)} \right) \, .
\label{eq:NLO_DECAY_RATE_FORMULA}
\end{align}
Here,  $d\phi_4, d\phi_5$ and $d\phi^{(\rm dip)}$
are the 4-particle, 5-particle, and dipole phase-space measures, respectively,
 $\delta{\cal M}_{\rm CT}$ denotes, schematically, the dipole
 subtraction counterterms for the two real radiation processes
 \eqref{tlnujrea},
 and 
\[ F_4(Y) = \Theta(Y_{b,g}-Y) \, ,\qquad \tilde{F}_4({Y}) = \Theta(\tilde{Y}_{(ij),{l}}-Y) \, , \qquad F_5(Y)=\sum_{i,j\neq
  l}\Theta(Y_{i,j}-Y)\Theta(Y-Y_{(ij),l}) \]
denote jet functions. The quantity $Y_{(ij),l}$ is calculated from the momentum of the pseudo-jet 
 that consists  of partons $i$ and $j$, cf. Sec. \ref{setup}, whereas the quantity $\tilde{Y}_{({ij}),{l}}$ is calculated from the emitter and spectator momenta $\tilde{k}_{ij}$ and $\tilde{k}_l$, which are defined in terms of the 5-particle phase space $\phi_5$.
 The following formulae are given for conventional dimensional regularisation.

\subsection*{Un-integrated dipoles}

 The set of counterterms $\delta{\cal M}_{\rm CT}$  for the 
 real radiation processes \eqref{tlnujrea} can be constructed, using
 the emitter-spectator terminology of \cite{Catani:1996vz},
            with  so-called final-final and final-initial dipoles.
We denote the 4-momenta of the top-quark and of the $b$ quark from
  the $tWb$ vertex with $k_t$ and $k_b$, and those of the two gluons
  or the $q, {\bar q}$ in  \eqref{tlnujrea} by $k_1, k_2$.
 
The final-final dipoles required for \eqref{tlnujrea} can be obtained
from \cite{Catani:1996vz}:
\begin{align}
  \D{b\rightarrow bg_1,g_2}^{\lambda_1\lambda_2}     = &  \frac{-1}{2
    k_1\cdot k_b} 4\pi\alpha_s C_F {\mu}^{2\epsilon}
  \left[\delta^{\lambda_1\lambda_2}\left(
      \frac{2}{1-Z_{b1,2}(1-Y_{b1,2})}-1-Z_{b1,2}-\epsilon
      (1-Z_{b1,2}) \right) \right] \, , \nonumber \\
  \D{b\rightarrow bg_2,g_1}^{\lambda_1\lambda_2}     = &  \frac{-1}{2
    k_2\cdot k_b} 4\pi\alpha_s C_F {\mu}^{2\epsilon} \left[\delta^{\lambda_1\lambda_2}\left( \frac{2}{1-Z_{b2,1}(1-Y_{b2,1})}-1-Z_{b2,1}-\epsilon (1-Z_{b2,1}) \right) \right]  \, ,\nonumber \\
  \D{g\rightarrow gg,b}^{\rho_1\rho_2}        = &  \frac{-1}{2
    k_1\cdot k_2} 8\pi\alpha_s C_A {\mu}^{2\epsilon}
  \left[-g^{\rho_1\rho_2}\left(
      \frac{1}{1-Z_{12,b}(1-Y_{12,b})}+\frac{1}{1-(1-Z_{12,b})(1-Y_{12,b})}-2 \right) \right.
                      \nonumber \\
     & \left. \, + \,\frac{1-\epsilon}{k_1\cdot k_2}\Pi^{\rho_1}_{\rm FF}\Pi^{\rho_2}_{\rm FF} \right] \, , \nonumber \\
  \D{g\rightarrow q\bar{q},b }^{\rho_1\rho_2}  = &  \frac{-1}{2
    k_1\cdot k_2} 4\pi\alpha_s N_f {\mu}^{2\epsilon} \left[-g^{\rho_1\rho_2} - \frac{2}{k_1.k_2}\Pi^{\rho_1}_{\rm FF}\Pi^{\rho_2}_{\rm FF} \right]. 
\label{eq:DIP_FF}
\end{align}
Here $\epsilon=(4-d)/2$, $C_F=(N_c^2-1)/(2N_c)$, $C_A=N_c$, $N_f=5$ and
\begin{align}
 Y_{ij,l}=\frac{k_i\cdot k_j}{k_i\cdot k_j+(k_i+k_j)\cdot k_l}   \, , & \quad
Z_{ij,l}=\frac{k_i\cdot k_l}{(k_i+k_j)\cdot k_l} \, , 
 \quad  (l \neq i,j) \, , \nonumber \\
 \Pi^{\rho_1}_{\rm FF} = (1-Z_{12,b})\, k^{\rho_1}_1-Z_{12,b}\, k^{\rho_1}_2 \, . & \nonumber
\end{align}
The indices $\lambda_i$ and $\rho_i$ transform according to the spinor and vector representations of the Lorentz group, i.e.,  they 
 refer to the spin
  of the initial-state  quark and gluon, respectively, in \eqref{eq:DIP_FF}.

\bigskip

The final-initial dipoles for $t\to Wbg_1g_2$ were constructed in
\cite{Melnikov:2011ta,Melnikov:2011qx} (using, in part,
results of \cite{Campbell:2004ch}). They contain the eikonal terms
$\propto {m_t^2k_i\cdot k_j}/{(k_t\cdot k_i)^2}$ for
  canceling the soft singularities that arise from gluon radiation off the initial top-quark.
 The final-initial dipole for
$t\to Wbq {\bar q}$ can be constructed analogously. 
\begin{align}
  \D{b\rightarrow bg_1 }^{t\,\lambda_1\lambda_2}      = &  \frac{-1}{2
    k_1\cdot k_b} 4\pi\alpha_s C_F {\mu}^{2\epsilon}
  \left[\delta^{\lambda_1\lambda_2}\left(
      \frac{2}{1-Z_{b1}^{t}}-1-Z_{b1}^{t}-\epsilon Y_{b1}^{t}(1-Z_{b1}^{t}) -
      \frac{m_t^2k_1\cdot
 k_b}{(k_t\cdot k_1)^2} \right) \right], \nonumber \\
  \D{b\rightarrow bg_2 }^{t\,\lambda_1\lambda_2}      = &  \frac{-1}{2
    k_2\cdot k_b} 4\pi\alpha_s C_F {\mu}^{2\epsilon}
  \left[\delta^{\lambda_1\lambda_2}\left(
      \frac{2}{1-Z_{b2}^{t}}-1-Z_{b2}^{t}-\epsilon Y_{b2}^{t}(1-Z_{b2}^{t}) -
      \frac{m_t^2k_2\cdot
 k_b}{(k_t\cdot k_2)^2} \right) \right], \nonumber \\
  \D{g\rightarrow gg }^{t\,\rho_1\rho_2}        = &  \frac{-1}{2
    k_1\cdot k_2} 8\pi\alpha_s C_A {\mu}^{2\epsilon}
  \left[-g^{\rho_1\rho_2}\left(
      \frac{1-Z_{12}^{t}}{Z_{12}^{t}}+\frac{1-Z_{21}^{t}}{Z_{21}^{t}} -
      \frac{m_t^2k_1\cdot
 k_2}{2(k_t\cdot k_1)^2} - \frac{m_t^2k_1\cdot k_2}{2(k_t\cdot k_2)^2} \right) \right. \nonumber\\
     & \left. \,+ \,\frac{1-\epsilon}{k_1\cdot k_2}\Pi^{\rho_1}_{\rm FI}\Pi^{\rho_2}_{\rm FI} \right], \nonumber \\
  \D{g\rightarrow q\bar{q}}^{t\,\rho_1\rho_2}  = &  \frac{-1}{2
    k_1\cdot k_2} 4\pi\alpha_s N_f {\mu}^{2\epsilon}
  \left[-g^{\rho_1\rho_2} - \frac{2}{k_1.k_2}\Pi^{\rho_1}_{\rm
      FI}\Pi^{\rho_2}_{\rm FI} \right] \, .
\label{eq:DIP_FI}
\end{align}
Here
\[ Z_{ij}^{t}=\frac{2 \, k_t.k_i}{m_t^2 \, (1-r_{ij}^2)} \, ,  \quad Y_{ij}^{t}=\frac{2 \, k_i.k_j} {m_t^2 \, (1-r_{ij})^2}\, ,
\quad r_{ij} = \frac{(k_t-k_i-k_j)^2}{m_t^2} \, . \]
 In this case the vector $\Pi^{\rho}_{\rm FI}$ takes a more
 complicated form. For the sake of brevity we refer to eq. (20) of \cite{Melnikov:2011qx}.

\subsection*{Integrated dipoles}

For the analytical 
       integration of
  the final-final dipoles over the respective subspaces  we use a phase-space splitting of the form

\begin{align}
 d\phi_5(k_i,k_j,k_l,\cdots) & =  d\phi_4(\tilde{k}_{ij},\tilde{k}_{l},\cdots)\times d\phi_{ij,l}^{\rm (dip.)}(Y_{ij,l},Z_{ij,l}) \nonumber,
\end{align}
such that one can, in the end, identify $d\phi_4(\tilde{k}_{ij},\tilde{k}_{l},\cdots)$ with the four-particle
 phase space of the Born matrix elements or of the virtual corrections.

The momenta of the emitter $\tilde{k}_{ij}$ and the 
 spectator $\tilde{k}_{l}$ are constructed according to \cite{Catani:1996vz} from the soft/collinear pair $k_i$, $k_j$ and another parton momentum $k_l$, whereas all remaining momenta are unaffected. In our case the emitter and spectator is either a b-quark or a gluon, i.e. $\tilde{k}_{ij}=\tilde{k}_{b/g}$ and $\tilde{k}_{l}=\tilde{k}_{g/b}$.

The phase space of the final-final dipoles can then be parameterized as

\begin{align}
 d\phi_{ij,l}^{\rm (dip.)}(Y_{ij,l},Z_{ij,l}) & =  \left[ \frac{\left(2\tilde{k}_{b}\cdot \tilde{k}_g\right)^{1-\epsilon} [d\Omega_{(d-3)}]}{16\pi^2 (2\pi)^{1-2\epsilon}} \right. \nonumber \\
& \left. \qquad \times \, \frac{\Theta\left(Z_{ij,l}(1-Z_{ij,l})\right) \; dZ_{ij,l}}{\left( Z_{ij,l}(1-Z_{ij,l}) \right)^{\epsilon}} \,  \frac{\Theta\left(Y_{ij,l}(1-Y_{ij,l})\right) \; dY_{ij,l}}{\left( 1-Y_{ij,l} \right)^{2\epsilon-1}Y_{ij,l}^{\epsilon}} \right].
\end{align}

The double index $ij$ labels the soft/collinear pair and the index $l$ refers to the momentum of  the remaining final-state parton.

Integration of \eqref{eq:DIP_FF} over the dipole phase space yields
\begin{align}
{D}_{b\rightarrow bg_1,g_2}^{\lambda_1\lambda_2}     = &  \frac{\alpha_s}{4\pi} \frac{-C_F}{\Gamma(1-\epsilon)} \left(\frac{\tilde{\mu}^2}{\tilde{S}_{bg}}\right)^{\epsilon} \left[\delta^{\lambda_1\lambda_2}\left( \frac{1}{\epsilon^2} + \frac{3}{2\epsilon}+5-\frac{\pi^2}{2} + {\cal O}(\epsilon) \right) \right]  \, ,\nonumber \\
{D}_{b\rightarrow bg_2,g_1}^{\lambda_1\lambda_2}     = &  \frac{\alpha_s}{4\pi} \frac{-C_F}{\Gamma(1-\epsilon)} \left(\frac{\tilde{\mu}^2}{\tilde{S}_{bg}}\right)^{\epsilon} \left[\delta^{\lambda_1\lambda_2}\left( \frac{1}{\epsilon^2} + \frac{3}{2\epsilon}+5-\frac{\pi^2}{2} + {\cal O}(\epsilon)  \right) \right]  \, ,\nonumber \\
{D}_{g\rightarrow gg,b}^{\rho_1\rho_2}        = &
\frac{\alpha_s}{4\pi} \frac{-2C_A}{\Gamma(1-\epsilon)}
\left(\frac{\tilde{\mu}^2}{\tilde{S}_{bg}}\right)^{\epsilon}
\left[-g^{\rho_1\rho_2}\left( \frac{1}{\epsilon^2} +
    \frac{11}{6\epsilon}+\frac{50}{9}-\frac{\pi^2}{2} + {\cal
      O}(\epsilon)  \right) \right]  \, , \nonumber \\
{D}_{g\rightarrow q\bar{q},b}^{\rho_1\rho_2}  = &  \frac{\alpha_s}{4\pi} \frac{-N_f}{\Gamma(1-\epsilon)} \left(\frac{\tilde{\mu}^2}{\tilde{S}_{bg}}\right)^{\epsilon} \left[-g^{\rho_1\rho_2}\left( -\frac{2}{3\epsilon}-\frac{16}{9} + {\cal O}(\epsilon)  \right) \right]  \, .
\label{eq:DIP_FF_INT}
\end{align}
Here $\tilde{\mu}^2=4\pi\mu^2$ and $\tilde{S}_{bg}=(\tilde{k}_b+\tilde{k}_g)^2/m_t^2 = {2\tilde{k}_b\cdot\tilde{k}_g}/{m_t^2}$.

\bigskip

In the case of the final-initial dipoles the phase-space splitting takes a slightly different form:

\begin{align}
 d\phi_5(k_i,k_j,R) & =  d\phi_4(\tilde{k}_{ij},\tilde{R})\times d\phi_{ij}^{t \;\rm (dip.)}(Y^{t}_{ij},Z^{t}_{ij}) \nonumber \, .
\end{align}

Again $d\phi_4(\tilde{k}_{ij},\tilde{R})$ can be identified with the phase space of the Born matrix elements or of
 the virtual corrections. Here, the phase-space mapping $5 \rightarrow 4$  affects, besides the soft/collinear pair, also 
 all other final state momenta, denoted by $R$. 

This procedure, as well as the dipole phase-space parameterization,

\begin{align}
 d\phi_{ij}^{\rm (dip.)}(Y_{ij}^{t},Z_{ij}^{t}) & =   \left[ \frac{\left(m_t^2\right)^{1-\epsilon} [d\Omega_{(d-3)}]}{16\pi^2 (2\pi)^{1-2\epsilon}} (1-r_{ij})^2\left(\frac{1+r_{ij}}{1-r_{ij}}\right)^{2\epsilon} \right.  \nonumber \\
            &    \left.\qquad  \times  \, \frac{\Theta\left(Z_{ij}^{t}(1-Z_{ij}^{t})\right) \; dZ_{ij}^{t}}{\left( Z_{ij}^{t} + r_{ij}^2(1-Z_{ij}^{t}) \right)^{\epsilon}} \, \frac{\Theta\left(Y_{ij}^{t}(Y_{\rm max}-Y_{ij}^{t})\right) \; dY_{ij}^{t}}{\left( Y_{\rm max}-Y_{ij}^{t} \right)^{\epsilon}\left(Y_{ij}^{t}\right)^{\epsilon}}
\right]. \nonumber
\end{align}

is adapted from \cite{Campbell:2004ch}, \cite{Melnikov:2011qx}.

The boundary of the $Y_{ij}^{t}$ integration is 
\[Y_{\rm max} = \frac{(1+r_{ij})^2Z_{ij}^{t}(1-Z_{ij}^{t})}{ (1-Z_{ij}^{t}) + r_{ij}^2 Z_{ij}^{t}} \, .\]

 Integration of  \eqref{eq:DIP_FI}  over the dipole phase space
 yields hypergeometric functions $_{2}{\rm F}_{1}$, which we expanded  in powers of
 $\epsilon$ using  the package HPL 2.0 \cite{Maitre:2005uu}. We obtain
\begin{align}
{D}_{b\rightarrow bg_1}^{t\,\lambda_1\lambda_2}    & =
\frac{\alpha_s}{4\pi} \frac{-C_F}{\Gamma(1-\epsilon)}
\tilde{\mu}^{2\epsilon} \left[\delta^{\lambda_1\lambda_2}\left(
    \frac{1}{\epsilon^2} + \frac{5-4\ln(1-\tilde{T}_b)}{2\epsilon}+F_{bg} +
    {\cal O}(\epsilon) \right) \right] \, ,  \label{eq:DIP_FI_INT1} \\
{D}_{b\rightarrow bg_2}^{t\,\lambda_1\lambda_2}    & =   \frac{\alpha_s}{4\pi} \frac{-C_F}{\Gamma(1-\epsilon)} \tilde{\mu}^{2\epsilon}  \left[\delta^{\lambda_1\lambda_2}\left( \frac{1}{\epsilon^2} + \frac{5-4\ln(1-\tilde{T}_b)}{2\epsilon}+F_{bg} + {\cal O}(\epsilon)  \right) \right] \, ,  \label{eq:DIP_FI_INT2} \\
{D}_{g\rightarrow gg}^{t\,\rho_1\rho_2}      & =   \frac{\alpha_s}{4\pi} \frac{-2C_A}{\Gamma(1-\epsilon)} \tilde{\mu}^{2\epsilon}  \left[-g^{\rho_1\rho_2}\left( \frac{1}{\epsilon^2} + \frac{17-12\ln(1-\tilde{T}_g)}{6\epsilon}+F_{gg}+ {\cal O}(\epsilon)  \right) \right]  \, , \label{eq:DIP_FI_INT3} \\
{D}_{g\rightarrow q\bar{q}}^{t\,\rho_1\rho_2} & =
\frac{\alpha_s}{4\pi} \frac{-N_f}{\Gamma(1-\epsilon)}
\tilde{\mu}^{2\epsilon}  \left[-g^{\rho_1\rho_2}\left(
    -\frac{2}{3\epsilon} + F_{q\bar{q}} + {\cal O}(\epsilon)  \right)
\right] \, ,
\label{eq:DIP_FI_INT4}
\end{align}
 where 
\begin{align}
F_{bg}      & =  - \ln\left( \tilde{T}_b \right)\, \frac{\tilde{T}_b \, \left(6 - 7 \, \tilde{T}_b\right)}{2 \, \left( 1-\tilde{T}_b \right)^{2} } + \frac{27-25\, \tilde{T}_b}{4\,\left(1-\,\tilde{T}_b\right)}   \nonumber \\
           &    + 2\,{\rm Li_2}(1-\tilde{T}_b)  - \frac{5}{6} \pi^2 - 5\,\ln\left( 1-\tilde{T}_b \right) + 2\, \ln^2\left( 1-\tilde{T}_b \right) , \nonumber \\
F_{gg}      & =   - \ln\left(\tilde{T}_g\right) \, \frac{ \tilde{T}_g \left(
    24-84\,\tilde{T}_g+134\,{\tilde{T}_g}^{2}-91\,\tilde{T}_g^3 +23\,\tilde{T}_g^4 \right)}{6\left( 1-\tilde{T}_g \right)^{5}} \nonumber \\
           &     + \, \frac{901-3694\,\tilde{T}_g+5326\,\tilde{T}_g^2-3534\,\tilde{T}_g^3+881\,\tilde{T}_g^4 }{ 120 \left( 1-\tilde{T}_g \right)^{4} } \nonumber \\
           &     + 2{\rm Li_2}(1-\tilde{T}_g) - \frac{5}{6} \pi^2 - \frac{17}{3}\, \ln\left(1-\tilde{T}_g\right) + 2\, \ln^2\left(1-\tilde{T}_g\right) ,   \nonumber \\
F_{q\bar{q}} & =   \ln\left( \tilde{T}_g \right) \,\frac{ \tilde{T}_g \left( 3+8\,\tilde{T}_g^2-7\,\tilde{T}_g^3+2\,\tilde{T}_g^4 \right)}{3\, \left( 1-\tilde{T}_g \right) ^{5}}  + \frac{4}{3}\, \ln  \left( 1-\tilde{T}_g \right)\nonumber \\
           &     - \frac{  \left( 101-494\,\tilde{T}_g+526\,\tilde{T}_g^2-334\,\tilde{T}_g^3+81\,\tilde{T}_g^4 \right) }{ 60 \left( 1-\tilde{T}_g \right) ^{4}}  . 
\label{eq:FIN_PARTS}
\end{align}
Here $\tilde{T}_b=(k_t-\tilde{k}_b)^2/m_t^2,$  and $\tilde{T}_g=(k_t-\tilde{k}_g)^2/m_t^2.$
Eq. \eqref{eq:DIP_FI_INT1} agrees with the result of
\cite{Melnikov:2011ta}, and eqs. 
\eqref{eq:DIP_FI_INT2} and
 \eqref{eq:DIP_FI_INT3} with those of \cite{Melnikov:2011qx}.



\end{document}